\documentclass[prd,aps,floats,eqsecnum,nofootinbib]{revtex4}
\usepackage{amsmath,amssymb,verbatim,epsfig,rotating}
\newcommand{\be}{\begin{equation}}
\newcommand{\ee}{\end{equation}}
\newcommand{\bea}{\begin{eqnarray}}
\newcommand{\eea}{\end{eqnarray}}
\begin{document}
\title{\bf The Cluster Expansion for the Self-Gravitating gas and the
  Thermodynamic Limit}
\author{{\bf  H. J. de Vega$^{(a)}$}}
\author{ {\bf N. G. S\'anchez$^{(b)}$}}
\affiliation{$(a)$  Laboratoire de Physique Th\'eorique et Hautes Energies, \\
Universit\'e Paris VI et VII, Tour 16, 1er \'etage, 4, Place Jussieu, \\
75252 Paris, cedex 05, FRANCE. Laboratoire Associ\'e au CNRS UMR 7589.}
\affiliation{(b) Observatoire de Paris,  LERMA,
\\ 61, Avenue de l'Observatoire,
75014 Paris,  FRANCE. \\
Laboratoire Associ\'e au CNRS UMR 8112.}
\begin{abstract}
We develop the cluster expansion and the Mayer expansion for the
self-gravitating thermal gas and prove the existence and stability of
the thermodynamic limit  $ N, V \to \infty $ with $ N/V^{\frac13} $
fixed. The essential (dimensionless) variable is here $ \eta \equiv
\frac{G \, m^2 N}{V^{\frac13} \; T} $ (which is kept fixed in the
thermodynamic limit). We succeed in this way to obtain the expansion of
the grand canonical partition function in powers of the fugacity. The
corresponding  cluster coefficients behave  in the thermodynamic limit as $
\left(\frac{\eta}{N}\right)^{j-1} \; c_j $ where $ c_j $ are pure
numbers. They are expressed as integrals associated to tree cluster
diagrams. A bilinear recurrence relation for the  coefficients $ c_j $
is obtained from the mean field equations in the Abel's form. 
In this way the
large $ j $ behaviour of the $ c_j $ is calculated. This large $ j $
behaviour provides the position of the nearest singularity which
corresponds to the critical point (collapse) of the self-gravitating
gas in the grand canonical ensemble. Finally, we discuss why other
attempts to define a thermodynamic limit for the self-gravitating gas
fail. 
\end{abstract}
\date{\today}
\maketitle
\tableofcontents
\section{Introduction}

The self-gravitating gas has been the subject of attention since many
years\cite{hidro,nos1,nos2,otros,dospar}. 
In refs.\cite{nos1,nos2} we recently investigated the self-gravitating thermal
gas using Monte Carlo simulations, mean field methods and low density
expansions\cite{nos1,nos2}. We have shown that the system possess a
well defined infinite volume limit in the grand canonical (GC), the
canonical (C) and microcanonical (MC) ensembles when $ N, V \to \infty
$ keeping $ N/V^{\frac13} $ fixed. A relevant variable here is
dimensionless ratio
\be\label{etadef}
\eta \equiv \frac{G \, m^2 N}{V^{\frac13} \; T} \; ,
\ee
which is kept fixed in the $ N, V \to \infty$ limit. All physical
quantities per particle turn out to be functions of the single
variable $ \eta $ and are well defined and {\bf finite} in the
thermodynamic limit $ N, V \to \infty$ with $ \eta $ fixed.

In this paper we develop the cluster expansion and  Mayer's approach to
the self-gravitating thermal gas and provide a rigorous demonstration
of the existence of the thermodynamic limit $ N, V \to \infty$ with $
\eta $ fixed. 

The cluster expansion is a powerful tool allowing to express the
partition function in a power series of the density for short range
interactions \cite{hill}. We apply and adapt this method (the Mayer expansion)
which is purely combinatorial to the self-gravitating gas. We succeed
in this way to obtain the coordinate partition function $ Q_N $ as a
power series in  the thermodynamic limit. This is derived by
generalizing the saddle point method used in ref.\cite{hill}. The expansion is
obtained in terms of the coefficients $ c_n $ which are pure
numbers. More explicitly, 
\be\label{pci}
\frac{1}{N} \; \log Q_N(\eta) \buildrel{N \gg 1}\over= \frac{1}{\eta}
\; g(\eta \; t_{\eta} ) - \log t_{\eta} - 1 + {\cal
  O}\left(\frac{1}{N}\right) \; . 
\ee
where
\be \label{gind}
g(x) \equiv \sum_{j=1}^{+\infty} c_j \; x^j \; .
\ee
and $t_{\eta}$ is the solution of the equation,
$$
t_{\eta} \; g'(t_{\eta} \; \eta ) = 1 \quad , \quad \mbox{i. e.} \quad
\sum_{j=1}^{+\infty} j \; c_j \; (t_{\eta} \;\eta )^j = \eta \; .
$$
Moreover, $ -\frac{1}{N} \; \log Q_N(\eta)$ is the free energy of the
self-gravitating gas minus the free energy of an ideal gas divided by
$NT$. The series for $ g(\eta \; t_{\eta} ) $ in
eqs.(\ref{pci})-(\ref{gind}) is therefore a high temperature or low
density expansion [see eq.(\ref{etadef})].

\bigskip				

The coefficients $ c_n $ can be expressed in the thermodynamic limit
as a sum of $3n$-uple integrals associated to tree cluster
diagrams. Loop cluster diagrams are subdominant for $ N \to \infty $.
The coefficients $ c_n $ only depend on the geometry of the box. 
The first  $ c_n $ are obtained by explicit evaluation of the cluster
integrals. We have for the sphere,
$$ 
c_1 = 1 \quad , \quad c_2^{sphere} = \frac35 \,
\left(\frac{4\pi}{3}\right)^{1/3}\quad ,  \quad 
c_3^{sphere} =\frac{51}{70}\, \left(\frac{4\pi}{3}\right)^{2/3}
\quad ,  \quad c_4^{sphere} = \frac{4\pi}{3} \;\frac{373}{315} \; . 
$$
Moreover, we use the connection with the mean field approach to obtain
a nonlinear recurrence relation for the coefficients $ c_n^{sphere} $: 
$$
c^R_n = \frac{1}{(2n+1)(n-1)} \sum_{s=1}^{n-1}c^R_s \; c^R_{n-s} \; s^2 \;
(2 \, n-2 \, s +1) \quad \mbox{for}  \quad n \geq 2 \; ,
$$
where 
$$
c_n^{sphere}= \left(\frac{4\pi}{3}\right)^{\frac{n-1}{3} } \; c^R_n \; .
$$
(The label $R$ stands for the spherical geometry).
This allows to compute systematically all $ c_n^{R} $ and to find
their large $n$ behaviour as
$$
c^R_n  \buildrel{n \gg 1}\over=  0.309360\ldots
\; \frac{n^{-\frac52}}{( u_{GC} )^n}
\left[1 + {\cal O}\left(\frac{1}{n}\right) \right] \; . 
$$
Here, $ u_{GC}=0.30034\ldots $ is the radius of convergence of the
series eq.(\ref{gind}). Therefore, $ u=u_{GC} $ is the nearest singularity of
$ g_R(u) $ in the $u$-plane.

\bigskip

We show below that the function $ g $ is related to the partition
function $ {\cal Z}_{GC}(\eta,z) $ in the grand canonical ensemble by
$$
\log {\cal Z}_{GC}(\eta,z) = \frac{N}{\eta} \; g(\eta \; t_{\eta}) \; ,
$$
where
$$
{\cal Z}_{GC}(\eta,z) = \sum_{N=0}^{+\infty} \frac{Q_N(\eta)}{ N! }\;\left[
  \left(\frac{mT}{2\pi}\right)^{\frac32} \; V \; z \right]^N \; .
$$
In addition, the fugacity ($ z = e^{\frac{\mu}{T}} $) turns out to be
given by $ z = t_{\eta} \;  e^{\frac{\mu_0}{T}} $, where $ \mu $ and
$ \mu_0 $ are the chemical potentials for the self-gravitating gas and
for the ideal gas, respectively.

\bigskip				

The point $ u_{GC}=0.30034\ldots $ where $ g(u) $ is singular
corresponds to the critical point (collapse) in the grand canonical
ensemble. Recall that the collapse point depends on the ensemble
considered \cite{nos1}.

\bigskip

In conclusion, our investigation here on the cluster expansion and the
Mayer expansion for the self-gravitating gas shows that:

\begin{itemize}

\item The self-gravitating gas admits a consistent thermodynamic limit
  $ N, V \to 
  \infty $ with $ \frac{N}{V^{\frac13}} $ fixed. In this limit,
  extensive thermodynamic quantities like energy, free energy, entropy
  are proportional to $N$. That is, in this limit the energy, free energy, and
  entropy {\bf per particle} are well defined and {\bf finite}.

\item The cluster expansion and the mean field approach provide the {\bf
  same} results in the thermodynamic limit  $ N, V \to 
  \infty $ with $ \frac{N}{V^{\frac13}} $ fixed.

\item The partition function needs a small short-distance cutoff $a$ in
  order to be well defined, the thermodynamic limit has a  {\bf finite limit}
  for $ a \to 0 $. In other words, the contributions to the partition
  function which diverge for $ a \to 0 $ are {\bf subdominant} for $ N \to
  \infty $ (see sec. II). For large $N$ and fixed cutoff $a$
  the potentially divergent contributions for $ a \to 0 $ are
  suppresed by a factor 
  at least $\eta^2/N^2$ compared with the dominant contribution for $ N \to
  \infty $. That is, the $ N, V =  \infty $ limit of the cutoff model
  (with $ \frac{N}{V^{\frac13}} $ fixed) has a {\bf finite} limit for
  $ a \to 0 $. Therefore, subdominant  corrections in $\eta/N$
  can always be neglected. Realistic models of the self-gravitating gas 
  (interstellar medium, galaxy distribution) require a small non-zero
  short distance cutoff since molecular  and atomic forces dominate
  over gravitational forces for short distances.

\end{itemize}

\section{The Cluster Expansion for the Self-Gravitating Gas}

We investigate in this section  the  self-gravitating gas in thermal
equilibrium at temperature $ T \equiv \beta^{-1} $. That is, we work
in the canonical ensemble where the
system of $ N $ particles is in contact with a thermal bath at
temperature $ T $. We assume the gas being  on a cubic box of side $ L $.

The partition function of the system can be written as
\begin{equation}\label{fp}
{\cal Z}_C(N,T) = \frac{1}{N !}\int\ldots \int
\prod_{l=1}^N\;\frac{d^3p_l\, d^3q_l}{(2\pi)^3} \; e^{- \beta H_N}\; ,
\end{equation}
where
\begin{equation}\label{hamic}
H_N = \sum_{l=1}^N\;\frac{p_l^2}{2m} - G \, m^2 \sum_{1\leq l < j\leq N}
\frac{1}{ |{\vec q}_l - {\vec q}_j|_A}\; .
\end{equation}
$G$ is Newton's gravitational constant.

\bigskip				

At short distances,  the particle interaction for the self-gravitating
gas in physical situations is not gravitational. Its
exact nature depends on the problem under consideration (opacity limit,
Van der Waals forces for molecules etc.). 
We shall just assume a repulsive short distance potential, that is,
\begin{equation}\label{defva}
v_A(|{\vec q}_l - {\vec q}_j|) = - \frac{1}{|{\vec q}_l - {\vec q}_j|_A } =
\left\{ \begin{array}{c}  - \frac{1}{|{\vec q}_l - {\vec q}_j|} \quad
\mbox{for} \; |{\vec q}_l - {\vec q}_j| \ge A \cr 
 \cr +\frac{1}{A} \quad \mbox{for}\; |{\vec q}_l - {\vec q}_j| \le A
\end{array} \right. 
\end{equation}
where $ A << L $ is the short distance cut-off. 

\bigskip				

The integrals over the momenta $p_l, \; (1 \leq l \leq N) $ in
eq.(\ref{fp}) can be computed immediately.

It is convenient to introduce the dimensionless variables $ {\vec r}_l
,\;  1\leq l \leq N $ making explicit the volume dependence  as
\begin{eqnarray}\label{variar}
{\vec q}_l &=& L \; {\vec r}_l \quad , \quad {\vec r}_l =(x_l,y_l,z_l)
\;, \cr \cr
0&\leq& x_l,y_l,z_l \leq 1\; .
\end{eqnarray}
That is, in the new coordinates the gas is inside a cube of unit volume.

The partition function takes now the form,
\begin{equation}\label{fp2}
{\cal Z}_C(N,T) = \frac{1}{N !}\left(\frac{m T L^2}{2\pi}\right)^{\frac{3N}2}
\; Q_N(\eta) \; ,
\end{equation}
where
\begin{equation}\label{QN}
 Q_N(\eta) \equiv \int_0^1\ldots \int_0^1
\prod_{l=1}^N d^3r_l\;\; e^{ \eta \; u({\vec r}_1,\ldots,{\vec r}_N)}\; ,
\end{equation}
$ \eta $ is the  dimensionless variable \cite{nos1}
\begin{eqnarray}\label{defeta}
\eta &\equiv& \frac{G \, m^2 N}{L \; T} \; ,
\end{eqnarray}
and $ u( {\vec r}_1, \ldots,{\vec r}_N) $ is defined by
\begin{equation}\label{defu}
u({\vec r}_1, \ldots, {\vec r}_N) \equiv \frac{1}{N}\sum_{1\leq l <
  j\leq N} \frac{1}{ |{\vec r}_l - {\vec r}_j|_a} \quad , \quad a
  \equiv A/L \ll 1 
\end{equation}
In this way {\bf all} dependence on the volume $ V = L^3 $ is buried
in the variable $ \eta $. 

\bigskip

The coordinate partition function $ Q_N(\eta) $ can be written as
\be\label{Qdef}
 Q_N(\eta) = \int_0^1\ldots \int_0^1
\prod_{l=1}^N d^3r_l\;\; \prod_{1\leq l < j\leq N} e^{\frac{\eta}{N \; 
 |{\vec r}_l - {\vec r}_j|_a}}\; .
\ee

We are now interested to expand $ Q_N(\eta) $ in powers of $
\frac{\eta}{N} $. In order to do this is convenient to define:
$$
f_{lj}  \equiv e^{\frac{\eta}{N \;  |{\vec r}_l - {\vec r}_j|_a}} - 1\; .
$$
For small $ \frac{\eta}{N} $ and fixed $ a>0 $ we have,
\be\label{flj}
f_{lj} = \frac{\eta}{N \;  |{\vec r}_l - {\vec r}_j|_a} + {\cal
  O}\left[\left(\frac{\eta}{N}\right)^2 \right]\; ,
\ee
All the integrals over $ r_l $ in eq.(\ref{QN}) are finite provided we
keep $ a>0 $. 

\bigskip

We can now multiply out the products of $f$'s in the coordinate
partition function $Q_N(\eta)$,
$$
\prod_{1\leq l < j\leq N} \left(1+f_{lj}\right) = 1 + \sum f_{ij} +
\sum f_{ij} \; f_{kl} + \ldots
$$
Thus, by introducing the $ f_{ij} $ functions the effects of
interparticle forces are better exhibited.

A systematic treatment of such sum of products can be found in
ref.\cite{hill}. The outcome is that a general term in the partition
function $Q_N(\eta)$ eq.(\ref{Qdef}),
can be factorized as the product of several integrals over the
coordinates $ {\vec r}_j $. Each integral corresponds to a `cluster'
of particles and is called $ b_j(\eta,N) $ where
\be \label{bj}
 b_j(\eta,N) = \frac{1}{j!} \int \int_0^1\ldots \int_0^1
\prod_{l=1}^j d^3r_l \; \; S_{1,2,\ldots,j}\; ,
\ee
with \cite{hill},
\bea \label{S}
&& S_j = 1 \cr \cr
&& S_{12} = f_{12} \cr \cr
&& S_{123} = f_{12} \; f_{23} +  f_{12} \; f_{13} + f_{13} \; f_{23} +
 f_{12} \; f_{13}\; f_{23} \cr \cr
&& S_{1234} = f_{12} \; f_{23} \; f_{34} + \mbox{nineteen \;
   permutations} +  \cr \cr
&&+ f_{12} \; f_{13} \; f_{24} \; f_{34} + \mbox{fourteen \;
   permutations}  +  \cr \cr
&& + f_{12} \; f_{13} \; f_{14} \; f_{24} \; f_{34} +
 \mbox{five \;    permutations} +  f_{12} \; f_{13} \; f_{14} \;
 f_{23} \; f_{24} \; f_{34} \; .
\eea
The main result is that $ Q_N(\eta) $ can be expressed as the infinite
sum, the so called `cluster expansion' as:
\be\label{desQ}
Q_N(\eta) = N! \; \sum_{{\widetilde m} \; , \sum_{j=1}^N (j \; m_j) = N}
\; \prod_{j=1}^N  \frac{\left[b_j(\eta,N)\right]^{m_j}}{m_j !}\; ,
\ee
where 
$$ 
{\widetilde m} \equiv (m_1, \ldots , m_N) \; .
$$
It must be stressed that eqs.(\ref{bj})-(\ref{desQ}) are purely
combinatorial and they apply both for the short range interactions
considered in ref.\cite{hill} as well as the long range Newton
potential.

\bigskip

Since $ f_{lj} =  {\cal   O}\left(\frac{\eta}{N}\right) $ for large $N$, and 
$S_{1,2,\ldots,j}$ contains at least a product of $ j $ factors $
f_{il} $, we see that
$$
S_{1,2,\ldots,j} =  {\cal   O}\left[\left(\frac{\eta}{N}\right)^{j-1}
  \right] \quad  ,  \quad \mbox{for}\quad N \gg j \; .
$$
Therefore, the cluster integrals eq.(\ref{bj}) take the form
\be\label{basi}
 b_j(\eta,N) = \left(\frac{\eta}{N}\right)^{j-1} \; c_j 
\left[ 1 + {\cal   O}\left(\frac{\eta}{N}\right)\right] \quad  ,
\quad \mbox{for}\quad N \gg j \; . 
\ee
Here the coefficients $c_j$ are positive numbers which only depend
on the geometry of the box. As shown in fig. 1 the dominant terms for
large $N$ are cluster diagrams with a tree structure. In the large $N$
limit, cluster diagrams with a loop structure as the last term in
eq.(\ref{S}) are subdominant.

\bigskip

From eqs.(\ref{flj})-(\ref{S}) and (\ref{basi}) we find,
\bea\label{c123}
&&c_1 = 1  \; ,\cr \cr
&&c_2 = \frac12 \int_0^1 \int_0^1 \frac{d^3 r_1 \; d^3 r_2}{ |{\vec r}_1
  - {\vec r}_2|}\; , \cr \cr
&&c_3= \frac12 \int_0^1 \int_0^1 \int_0^1 \frac{d^3 r_1 \; d^3 r_2 \;
d^3 r_3}{ |{\vec r}_1   - {\vec r}_2| \; |{\vec r}_2   - {\vec r}_3|}\; .
\eea
For a sphere of unit volume we find \cite{nos1}
\be \label{esfera}
c_2^{sphere} = \frac35 \, \left(\frac{4\pi}{3}\right)^{1/3} =
0.967195171\ldots \quad ,  \quad 
c_3^{sphere} =\frac{51}{70}\,
\left(\frac{4\pi}{3}\right)^{2/3}=1.893206013\ldots   
\ee
For the cubic geometry chosen, it takes the value \cite{nos1}
$$
c_2^{cube} = 4 \;  \int_0^1 (1-x) \, dx \int_0^1 (1-y)\, dy\int_0^1 
\frac{(1-z)\, dz}{\sqrt{x^2 + y^2 + z^2}} =0.94116\ldots  
$$
Furthermore, we find for the coefficient $ c_4 $,
\be\label{c4}
c_4= \frac12 \int_0^1 \int_0^1 \int_0^1 \int_0^1 \frac{d^3 r_1 \; d^3 r_2 \;
d^3 r_3 \; d^3 r_4}{ |{\vec r}_1   - {\vec r}_2| \; |{\vec r}_2   -
  {\vec r}_3| \; |{\vec r}_3   - {\vec r}_4|} + \frac16 
\int_0^1 \int_0^1 \int_0^1 \int_0^1 \frac{d^3 r_1 \; d^3 r_2 \;
d^3 r_3 \; d^3 r_4}{ |{\vec r}_1   - {\vec r}_2| \; |{\vec r}_2   -
  {\vec r}_3| \; |{\vec r}_2   - {\vec r}_4|}\; . 
\ee
The combinatorial factors $ \frac12 $ and $ \frac16 $ take into
account the symmetries of the respective cluster diagrams.

\bigskip

Inserting here the expansion in spherical harmonics,
$$
\frac{1}{|{\vec r}   - {\vec r}'|} = 4\pi\sum_{l=0}^{\infty}
\frac{1}{2l+1} \; \frac{r_<^l}{r_>^{l+1}} \; \sum_{m=-l}^{m=+l} {\bar
  Y}_{lm}({\hat r}') \; Y_{lm}({\hat r}) \; ,
$$
where $ r_> \equiv \mbox{max}(r,r') $ and  $ r_< \equiv \mbox{min}(r,r') $ ,
the angular integrals and then the radial integrals can be performed
with the result, 
$$
\int_0^1 \int_0^1 \int_0^1 \int_0^1 \frac{d^3 r_1 \; d^3 r_2 \;
d^3 r_3 \; d^3 r_4}{ |{\vec r}_1   - {\vec r}_2| \; |{\vec r}_2   -
  {\vec r}_3| \; |{\vec r}_3   - {\vec r}_4|} =(4\pi)^4 \; \int_0^b
\ldots \int_0^b \frac{\prod_{l=1}^{4} r_l^2 \; dr_l}{ r^>_{1,2} \;
  r^>_{2,3}  \; r^>_{3,4} } =\frac{4\pi}{3} \; \frac{62}{35} \; .
$$
and
$$
\int_0^1 \int_0^1 \int_0^1 \int_0^1 \frac{d^3 r_1 \; d^3 r_2 \;
d^3 r_3 \; d^3 r_4}{ |{\vec r}_1   - {\vec r}_2| \; |{\vec r}_2   -
  {\vec r}_3| \; |{\vec r}_2   - {\vec r}_4|}= (4\pi)^4 \; \int_0^b
\ldots \int_0^b \frac{\prod_{l=1}^{4} r_l^2 
  \; dr_l}{ r^>_{1,2} \; r^>_{2,3}  \; r^>_{2,4} } =
\frac{4\pi}{3} \; \frac{188}{105} \; .
$$
Therefore,
\be\label{c4r}
c_4^{sphere} = \frac{4\pi}{3} \;\frac{373}{315} \; .
\ee
Notice that all these `tree' cluster diagrams eqs.(\ref{c123}) and
(\ref{c4}) have a {\bf finite limit} for zero cutoff $ a $.

\bigskip

Divergent pieces for $ a \to 0 $ are subdominant as $
\frac{\eta^2}{N^2} $ for large $N$. The leading divergent contribution for 
$ a \to 0 $ to $ Q_N(\eta) $ to the nth. order in $ \eta $ takes the
form\cite{nos1} 
\be\label{subd}
\frac{ \eta^n}{ n! \; N^n } \frac12 N(N-1) \int 
\frac{ d^3r_1 \; d^3r_2}{|{\vec r}_1 - {\vec r}_2|_a^n } 
\buildrel{a \to 0}\over\sim 
\left\{ \begin{array}{l} \frac{ \eta^n}{ n! \; N^{n-2}}\;  a^{3-n}
\quad \mbox{for} \quad  n > 3 \; , \\
 \\ \frac{ \eta^3}{ N} \log a\quad \mbox{for}\quad  n=3 \; .
\end{array}    \right.
\ee
This gives for the physical quantities (see next section)
contributions of the order
$$
\frac{ \eta^n}{ n! \; N^{n-1}}\;  a^{3-n} \quad \mbox{for} \; n> 3
\quad \mbox{and} \quad   \frac{ \eta^3}{ N^2} \log a\quad \mbox{for}\quad  n=3
\; .
$$
These contributions are clearly negligible in the $ N \to \infty $
limit with fixed short-distance cutoff.

\section{The Mayer Expansion for the self-gravitating gas}

It is convenient to consider the generating function \cite{hill},
\be\label{Xb}
X(\eta,z)\equiv\sum_{j=1}^{+\infty}  b_j(\eta,N) \; z^j\; ,
\ee
It can be shown that \cite{hill}
\be\label{GFP}
\sum_{N=0}^{+\infty} \frac{Q_N(\eta)}{ N! } \; z^N = e^{X(\eta,z)} \; ,
\ee
where $z$ is an auxiliary variable whose physical meaning (the
fugacity) will appear in sec. IV.

\bigskip

Hence, we can compute the coefficients $ Q_N(\eta) $ from
eq.(\ref{GFP}) by contour integration,
$$
\frac{Q_N(\eta)}{ N! } = \oint \frac{dz}{2\pi i} \;
\frac{e^{X(\eta,z)}}{z^{N+1}}\; .
$$
We choose as contour a cercle of radius $r$,
$$
z = r \; e^{i \theta} \quad , \quad 0 \leq \theta \leq 2 \pi\; .
$$
Integrating over $ \theta $ yields,
$$
\frac{Q_N(\eta)}{ N! } = \int_{-\pi}^{+\pi} \frac{d\theta}{2\pi} \; 
e^{X(\eta, r \; e^{i \theta}) - N \; \log r -i \, N \; \theta} \; .
$$
For large $N$, we can use eqs.(\ref{basi}) to express the $
b_j(\eta,N) $ in $X(\eta,z)$ [see eq.(\ref{Xb})] and we find,
\be\label{Xlim}
X(\eta,z) \buildrel{N \gg 1}\over= \frac{N}{\eta}\sum_{j=1}^{+\infty} c_j
\left(\frac{z \; \eta}{N}\right)^j = \frac{N}{\eta} \; 
g\left(\frac{z  \; \eta}{N}\right) \; ,
\ee
where
\be
g(x) \equiv \sum_{j=1}^{+\infty} c_j \; x^j \; .
\ee
Thus,
\be\label{inteta}
\frac{Q_N(\eta)}{N!} \buildrel{N \gg 1}\over= \int_{-\pi}^{+\pi}
\frac{d\theta}{2\pi} \; e^{N\left[ \frac{1}{\eta} \; g\left(\frac{r
 \; e^{i \theta} \; \eta}{N}\right)- \log r-i \; \theta\right]}  =
\int_{-\pi}^{+\pi} \frac{d\theta}{2\pi} \; e^{N \; \Psi(r,\theta)} \; .
\ee
We can now apply the steepest descent method to this integral for
large $N$ since the integrand has the structure $ e^{N \; \Psi(r,\theta)} $
where,
\bea
&&\Psi(r,\theta) = \frac{1}{\eta} \; g\left(\frac{r  \; e^{i \theta} \;
\eta}{N}\right)- \log r-i \; \theta \; , \cr \cr
&& \frac{\partial \Psi}{\partial\theta} (r,\theta) = i \; \frac{r \; e^{i
\theta}}{N} \; g'\left(\frac{r  \; e^{i \theta} \; \eta}{N}\right) - i
 \; ,  \cr \cr
&&\frac{\partial \Psi}{\partial r} (r,\theta) =\frac{e^{i\theta}}{N} \;
g'\left(\frac{r  \; e^{i \theta} \; \eta}{N}\right) -\frac{1}{r}\; .
\eea
The saddle point, solution of $ \frac{\partial \Psi}{\partial\theta}
(r,\theta) = 0 = \frac{\partial \Psi}{\partial r} (r,\theta) $ is thus
found at 
$$ 
(r,\theta)_{\mbox{saddle}} = (N \; t,0) \; ,
$$ 
where $t$ is $N$-independent and is a solution of the equation,
\be\label{tdeta}
t \; g'(t \; \eta ) = 1 \; .
\ee
That is, $ t $ is a function of $ \eta $ defined by the constraint
eq.(\ref{tdeta}) or more explicitly,
$$
\sum_{j=1}^{+\infty} j \; c_j \; (t \;\eta )^j = \eta \; .
$$

Choosing $ r = N \; t $ as integration path in eq.(\ref{inteta}) and
expanding the integrand around $ \theta = 0 $ yields,
$$
\frac{Q_N(\eta)}{ N! } \buildrel{N \gg 1}\over= e^{N \;\Psi(N \, t,0)}
\int_{-\infty}^{+\infty} \frac{d\theta}{2\pi} \; e^{-\frac12 \, N \;
  \theta^2 \; \xi(\eta) }=
\frac{e^{N \;\Psi(N \, t,0)}}{\sqrt{2 \, \pi \; N \;\xi(\eta)}}\left[
  1 + {\cal   O}\left(\frac{1}{N}\right)\right] \; , 
$$
where,
\bea
&& \Psi(N \, t,0) = \frac{1}{\eta} \; g\left(\eta \; t_{\eta}\right)-
\log [N \; t_{\eta}] \cr \cr
&&\xi(\eta)\equiv - \frac{\partial^2 \Psi}{\partial\theta^2}(Nt,0) =
  \sum_{j=1}^{+\infty} j^2 \; c_j \; t^j > 0  \; .
\eea
where $t_{\eta}$ is a function of $\eta$ defined by eq.(\ref{tdeta}).

\bigskip

Using now Stirling's formula for the $N!$ factorial we find,
\be\label{sti}
\frac{1}{N} \; \log Q_N(\eta) = \frac{1}{\eta} \; g(\eta \; t_{\eta} ) -
\log t_{\eta} - 1 + {\cal O}\left(\frac{1}{N}\right) \; .
\ee
Therefore, the free energy can be written as,
$$
\frac{F - F_0}{NT} = -\frac{1}{\eta} \; g(\eta \; t_{\eta} ) + \log
t_{\eta} + 1 \; ,
$$
where $F_0$ stands for the free energy of the ideal gas,
\begin{equation}\label{Fcero}
F_0 = -N T \log\left[\frac{ e V}{N}
  \left(\frac{mT}{2\pi}\right)^{\frac32}\right]\; . 
\end{equation}
The pressure (at the surface) follows from the thermodynamic relation
$$
p = - \left(\frac{ \partial F}{  \partial V}\right)_T \; ,
$$
we find 
\be\label{feta}
f(\eta) \equiv \frac{p V}{ N T} = \frac23 + \frac{g(\eta \; t_{\eta} )
}{3 \; \eta} \quad \mbox{and}  \quad \frac{1}{N} \; \log Q_N(\eta) = 3
\; \int_0^{\eta} dx \, \frac{1 - f(x)}{x}\; . 
\ee
The dimensionless ratio $\frac{p V}{ N T}$ was called $ f(\eta) $ in
refs.\cite{nos1,nos2}. We can express all physical quantities in terms
of the function $ f(\eta) $.
We find from eqs.(\ref{sti}) and (\ref{feta}),
\be \label{ident}
3 \; f(\eta) = 2 +  \frac{g(\eta \; t_{\eta} ) }{\eta}
\quad \mbox{and} \quad 
 -\frac{1}{\eta} \; g(\eta \; t_{\eta} ) + \log
t_{\eta} + 1 = - 3 \; \int_0^{\eta} dx \, \frac{1 - f(x)}{x}\; .
\ee
That is,
\be\label{tfeta}
\frac13 \log  t_{\eta} =  f(\eta) - 1 - \int_0^{\eta} dx \, \frac{1 -
  f(x)}{x}\; . 
\ee
We can solve eq.(\ref{tdeta}) for $t_{\eta}$ in powers of $\eta$ with the
result,
\bea
&&g(\eta \; t_{\eta}) = \eta \; t_{\eta} + c_2 \; (\eta \; t_{\eta})^2 +
c_3 \; (\eta \; t_{\eta})^3 +  {\cal O}(\eta^4) \; , \cr \cr
&& t_{\eta} = 1 - 2 \; c_2 \; \eta + \left[ 8 \;  (c_2)^2 - 3 \; c_3
  \right] \; \eta^2 + {\cal O}(\eta^3) \; .
\eea
and from eq.(\ref{ident}) we find for $f(\eta)$ and $g(\eta\; t_{\eta})$,
\bea
&&f(\eta) = 1 - \frac{c_2}{3} \; \eta + \frac{2}{3} \left[2 \;  (c_2)^2 -
  c_3\right] \; \eta^2 + {\cal O}(\eta^3) \; , \cr \cr
&& - 3 \; \int_0^{\eta} dx \, \frac{1 - f(x)}{x} = - c_2 \, \eta +
\left[2 \;  (c_2)^2 -  c_3\right] \; \eta^2 + {\cal O}(\eta^3) \cr \cr
&& g(\eta\; t_{\eta}) = \eta \left\{ 1 - c_2 \, \eta + 2 \; \left[2 \;
  (c_2)^2 -   c_3\right] \; \eta^2 \right\} + {\cal O}(\eta^4) \; .
\eea
It must be noticed that the function $ f(\eta) $ (and hence all
physical quantities) have the {\bf same}  expression whether we compute it
from the mean field approach \cite{nos1,nos2} or from the saddle point
eqs.(\ref{inteta}), (\ref{sti}) and (\ref{feta}). 

\section{The Grand Canonical Ensemble for the self-gravitating gas}

The definition of the variable $z$ through eq.(\ref{GFP}) suggests 
that $z$ is related to the fugacity. To be more precise, the grand
partition function is defined in terms of the canonical partition
function as,
\be\label{zgcano}
{\cal Z}_{GC}(\eta,z) = \sum_{N=0}^{+\infty} {\cal Z}_C(N,T) \; z^N
=\sum_{N=0}^{+\infty} \frac{Q_N(\eta)}{ N! }\;\left[
  \left(\frac{mT}{2\pi}\right)^{\frac32} \; V \; z \right]^N
\ee
where we used eq.(\ref{fp2}).

Therefore, $z$ must be multiplied by the ideal gas factor $
\left(\frac{mT}{2\pi}\right)^{\frac32} \; V $. 

\bigskip

At the saddle point we have after this renormalization 
\be\label{reno}
\left(\frac{mT}{2\pi}\right)^{\frac32} \; V \; z = N \; t_{\eta} \quad \mbox{and}
\quad X\left(\eta, \left(\frac{mT}{2\pi}\right)^{\frac32} \; V \; z\right)
= \log {\cal Z}_{GC}(\eta,z) \; .
\ee
Now, we know from ref.\cite{nos1} that the chemical potential $\mu$ takes
the form
\be\label{potq}
\frac{\mu - \mu_0}{T} = - 3 \; \int_0^{\eta} dx \, \frac{1 - f(x)}{x} -
3[1 - f(\eta)]
\ee
where $\mu_0$ stands for the chemical potential of the ideal gas,
\be\label{mu0}
\mu_0 = -T \, \log\left[\frac{V}{N}
\left(\frac{mT}{2\pi}\right)^{\frac32}\right] \; .
\ee 
From eq.(\ref{tfeta}) and (\ref{potq}) we have,
\be\label{logt}
\frac{\mu -\mu_0}{T} = \log t_{\eta} \; .
\ee
Therefore, from eqs.(\ref{reno})-(\ref{logt}),
\be\label{zt}
t_{\eta}=e^{-\frac{\mu_0}{T}} \; z \quad , \quad e^\frac{\mu}{T} = z \; ,
\ee 
that is, we can identify $ z $ with the fugacity. 

\bigskip

We have found in ref.\cite{nos1} that $ f(\eta) $ obeys in the
spherical case the first order differential equation of Abel's type,
\be\label{abel}
\eta^R[3 \; f(\eta^R)-1]f'(\eta^R)+[3f(\eta^R)-3+\eta^R] f(\eta^R) = 0 \; .
\ee
where the variable $\eta^R$ is defined as,
$$
\eta^R \equiv \eta \; \left(\frac{4\pi}{3}\right)^{1/3}  =
1.61199\ldots \; \eta \; .
$$
Using eqs.(\ref{abel}) and (\ref{tfeta}) yields,
\be \label{teta}
\log t_{\eta} = \log f(\eta^R) - \eta^R \quad , \quad \mbox{i. e.}\quad
t_{\eta}= e^{-\eta^R} \;  f(\eta^R) \; .
\ee
Eq.(\ref{zgcano}) provides the grand partition function. The
above results showing that $ {\cal Z}_{GC}(\eta,z) $ is dominated by
the canonical ensemble, together with  eq.(\ref{zt}) 
prove that the canonical and grand canonical ensembles are
equivalent in their common region of validity as stated in
refs.\cite{nos1,nos2}. 

\bigskip

In the thermodynamic limit,  $ {\cal   Z}_{GC}(\eta,z) $ is
given by eqs.(\ref{Xlim}) and (\ref{reno}) as
\be\label{zgcg}
\log {\cal Z}_{GC}(\eta,z) = \frac{N}{\eta} \; g\left(\frac{z  \; \eta}{N} \;
\left(\frac{mT}{2\pi}\right)^{\frac32} \; V\right)=\frac{N}{\eta} \;
g(\eta \; t_{\eta}) \; .
\ee
Using here eq.(\ref{ident}) gives,
$$
\log {\cal Z}_{GC}(\eta,z) = N [ 3 \; f(\eta) - 2 ]\; ,
$$
which exactly coincides with the expression found in ref.\cite{nos1}
for the grand partition function in the mean field approach. 

\section{Calculation of the Cluster coefficients $ {\bf {\Large c_n}}$ }

We compute in this section the cluster coefficients $ c_n $ for the
sphere by using a non-linear differential equation for the function $
g(u) $. 

We first find from eq.(\ref{tdeta})
\be\label{defgR}
u \, g'_R(u) = \eta^R \quad \mbox{with} \quad u \equiv t_{\eta} \;
\eta^R \quad \mbox{and} \quad g_R(u) \equiv
\left(\frac{4\pi}{3}\right)^{1/3} \; g(u) 
\ee
Then, combining eqs.(\ref{abel}) and (\ref{defgR}) yields 
the second order differential equation for $ g_R(u) $,
\be\label{eqdg}
\left[ u^2 \, g''_R(u) + u \,  g'_R(u) \right] \left[ 2 \, u \,
  g'_R(u) +  g_R(u) \right] - 2 \, u^2 \, g''_R(u) -  u \,  g'_R(u) +
g_R(u) = 0 \; .
\ee
Alternatively, if we choose $ \eta^R =  u \,  g'_R(u) $  as variable
[see eq.(\ref{defgR})], we find from eq.(\ref{abel}) the first order
non-linear differential equation 
$$
\left[ \eta^R + g_R \right] \frac{dg_R}{\;d\eta^R} + \eta^R
\left[2 \, \eta^R   +g_R -2 \right] = 0 \; .
$$
That is, the invariance of eq.(\ref{eqdg}) under the rescaling of the
variable $ u $ allows to reduce by one the order of the differential equation.

Eq.(\ref{defgR}) has as regular solution around $ u = 0 $,
\be\label{solg}
g_R(u) = \sum_{n=1}^{\infty} c^R_n \; u^n \quad \mbox{with} \quad
c_n^{sphere}= \left(\frac{4\pi}{3}\right)^{\frac{n-1}{3} } \; c^R_n \; ,
\ee
and we can choose $ c^R_1 = 1 $. 

\bigskip

By inserting eq.(\ref{solg}) into eq.(\ref{eqdg}) the following
 nonlinear recurrence relation for the  $ c^R_n
$ coefficients is obtained:
\be\label{relrec}
c^R_n = \frac{1}{(2n+1)(n-1)} \sum_{s=1}^{n-1}c^R_s \; c^R_{n-s} \; s^2 \;
(2 \, n-2 \, s +1) \quad \mbox{for}  \quad n \geq 2 \; .
\ee
We find,
$$
c^R_1=1\; , \; c^R_2 = \frac35 \; , \; c^R_3 = \frac{51}{70} \; , \; c^R_4 =
\frac{373}{315} \; ,  \; c^R_5 = \frac{14911}{6600} \; ,
\; c^R_6 =\frac{2047}{429} \; , \ldots
$$
in agreement with eqs. (\ref{c123}), (\ref{esfera}) and (\ref{c4r}).

\bigskip

By evaluating numerically the $ c^R_n $ from the
recurrence relation eq.(\ref{relrec}), we find for large $n$:
\be\label{gror}
c^R_n  \buildrel{n \gg 1}\over=  0.309360\ldots
\; \frac{n^{-\frac52}}{( u_{GC} )^n}
\left[1 + {\cal O}\left(\frac{1}{n}\right) \right] \; . 
\ee
Here, $ u_{GC}=0.30034\ldots $ is the radius of convergence of the
series eq.(\ref{solg}). Therefore, $u=u_{GC}$ is the nearest singularity of
$g_R(u)$ in the $u$-plane.

\bigskip

As we see from eq.(\ref{zgcg}), $g_R(u)$ gives the grand canonical
partition function as a function of $ u = t_{\eta} \; \eta^R$. Since $ u =
t_{\eta} \; \eta^R$ is proportional to the fugacity [see eq.(\ref{zt})], 
$u_{GC}$ must be related to the critical point of the selfgravitating gas in 
the grand canonical ensemble. Indeed, using the  critical value for $
\eta^R $ in the grand canonical ensemble \cite{nos1},  
$$ 
\eta^R_{GC} = 0.79735 \ldots \quad \mbox{and}\quad  f(\eta^R_{GC}) =
\frac{2}{3 \; \eta^R_{GC}} \; ,
$$
from eq.(\ref{teta}) we obtain, 
$$
u_{GC} = \eta^R_{GC} \; t_{\eta^R_{GC}} =
\eta^R_{GC}  \; e^{-\eta^R_{GC}} \; f(\eta^R_{GC}) = \frac23 \;
e^{-\eta^R_{GC}} = 0.30034\ldots \; ,
$$
in perfect agreement with the value for $u_{GC}$ in eq.(\ref{gror}). 
The large order behaviour of the expansion coefficients eq.(\ref{gror}) 
corresponds to a $ ( u_{GC} - u )^{\frac32} $ behaviour of the
function  $g_R(u)$. More precisely, for $ u \to  u_{GC} $ from
eq.(\ref{eqdg}) we find,
$$
g_R(u)\buildrel{u \to  u_{GC}}\over= 2(1-\eta^R_{GC}) +
\eta^R_{GC}\left(\frac{u}{ u_{GC}} - 1 \right) + \frac23 \;
  \sqrt{2-\eta^R_{GC}}\left(1-\frac{u}{ u_{GC}}
\right)^{\frac32} + {\cal O}\left[(u-u_{GC})^2\right] \; .
$$
Expanding this asymptotic behaviour in powers of $u$ yields,
$$
g_R(u)\buildrel{u \to  u_{GC}}\over= 2(1-\eta^R_{GC}) +
\eta^R_{GC}\left(\frac{u}{ u_{GC}} - 1 \right) + \frac12
\frac{\sqrt{2-\eta^R_{GC}}}{\pi} \sum_{k=0}^{\infty} \frac{\Gamma(k -
    \frac32)}{k! \;  u_{GC}^k} \; u^k \; .
$$
This implies for the coefficients $c^R_n$ the following large order behaviour,
$$
c^R_n  \buildrel{n \gg 1}\over= \frac12 \; \sqrt{\frac{2 -
    \eta^R_{GC}}{\pi}} \; \frac{n^{-\frac52}}{( u_{GC} )^n}
\left[1 + {\cal O}\left(\frac{1}{n}\right) \right] \; . 
$$
which exactly coincides with eq.(\ref{gror}) since,
$$
\frac12 \; \sqrt{\frac{2-\eta^R_{GC}}{\pi}} = 0.309360\ldots \; .
$$

\bigskip

In summary, the investigation here on the cluster expansion for the
selfgravitating gas shows:

\begin{itemize}

\item The selfgravitating gas admits a consistent thermodynamic limit
  $ N, V \to 
  \infty $ with $ \frac{N}{V^{\frac13}} $ fixed. In this limit,
  extensive thermodynamic quantities like energy, free energy, entropy
  are proportional to $N$. That is, in this limit the energy, free energy, and
  entropy {\bf per particle} are well defined and {\bf finite}.

\item The cluster expansion and the mean field approach provide the {\bf
  same} results in the thermodynamic limit  $ N, V \to 
  \infty $ with $ \frac{N}{V^{\frac13}} $ fixed.

\item The partition function needs a small short-distance cutoff $a$ in
  order to be well defined, the thermodynamic limit has a  {\bf finite limit}
  for $ a \to 0 $. In other words, the contributions to the partition
  function that diverge for $ a \to 0 $ are {\bf subdominant} for $ N \to
  \infty $ [see eq.(\ref{subd})]. That is, the $ N, V =  \infty $ limit of the
  cutoff model 
  (with $ \frac{N}{V^{\frac13}} $ fixed) has a {\bf finite} limit for
  $ a \to 0 $.  

\end{itemize}

\section{The stability of the thermodynamic limit {\bf ${\bf N,V=L^3 \to
  \infty}$} with ${\bf \frac{\bf N}{\bf L}}$ fixed (fixed ${\bf\eta}$) }

As shown in refs.\cite{nos1} there are two phases in the
selfgravitating gas: $ \eta < \eta_0 $ and $ \eta > \eta_0 $ where $
\eta_0 = 1.51024\ldots $ for spherical geometry and $ \eta_0 \simeq
1.515 $ for cubic geometry. For $ \eta > \eta_0 $ the selfgravitating
gas {\bf collapses} into a extremely dense phase with large and
negative pressure. For $ \eta < \eta_0 $ the selfgravitating gas is
perfectly {\bf stable}. The mean field applies in this gaseous phase
and coincides with the expansion in powers of $ \eta $ discussed in
the previous section. Such expansion converges within the gaseous
phase. The variable $ \eta $ is related to the Jeans' length of the
system as 
$$
\eta = 3 \; \left( \frac{L}{d_J}\right)^2 \quad , \quad L = V^{\frac13}
$$
where 
$$ 
d_J = \sqrt{\frac{3 \, T}{m}} \; \frac{1}{\sqrt{G\, m \,
    \rho}} \quad , \quad \rho \equiv \frac{N}{V} 
$$
The gas collapses in the canonical ensemble for $  \eta > \eta_0 $
which corresponds to $ L \gtrsim d_J $. This corresponds to the Jeans'
instability. 

The relevance of the ratio $\frac{G \, m^2}{ V^{\frac13} \; T} $ has
been noticed on dimensional grounds\cite{mongo}. However, the
dimensionality argument alone cannot single out the crucial factor $N$
in the variable $ \eta $. Notice that  $ \eta $ contains the ratio $
\frac{N}{V^{\frac13} } $ and not $ \frac{N}{V} $. Therefore, in the
thermodynamic limit 
$$ 
V \sim N^3  \quad \mbox{and}   \quad\rho = \frac{N}{V} \sim
\frac{1}{N^2} \to 0 \; .
$$ 
As $ N,V \to \infty , \; \eta $ is kept fixed
in the same way as the temperature $ T $. The energy, the product
$PV$, the free energy, the entropy are expressed as $N$ times
functions of $ \eta $. The chemical potential, the specific heats, the
compressibilities are just functions of $ \eta $.

\bigskip

In a recent e-print \cite{laliena} it was stated that the thermodynamic
functions for the selfgravitating gas diverge in the thermodynamic
limit $ N, V \to  \infty $ with $ \frac{N}{V^{\frac13}} $ fixed.

We show here below that the statements made in ref.\cite{laliena} have crucial
failures which invalidate the conclusions given in ref.\cite{laliena}.

Such statements in ref.\cite{laliena} are based in the inequality [eq.(30) in
ref.\cite{laliena}],
$$
\mathcal{Z}_\mathrm{C}  \geq
\frac{1}{N!}\,\beta^{-3N/2}\,\int_{V_0^N}\,
\prod_{i=1}^N d^3r_i\,\exp[-\beta \sum_{i<j}\phi_{ij}] \geq
\frac{V_0^N}{N!} \exp[-\beta \sum_{i<j} \langle \phi_{ij} \rangle_{V_0}] \, ,
$$
where
$$
\langle\phi_{ij}\rangle_{V_0}=\frac{1}{V_0^N}\,
\int_{V_0^N}\prod_{k=1}^N d^3r_k \, \phi_{ij} \; , \; \mbox{and} \;
\phi_{ij}\equiv -\frac{G}{|{\vec r}_i - {\vec r}_j|} \; ,
$$
and the last inequality follows from the property of the exponential 
function:
$\langle \exp(y)\rangle \geq \exp(\langle y\rangle)$.

\bigskip

To write such inequality the author of ref.\cite{laliena} considers $N$
 selfgravitating particles in a volume $ V = R^3 $ with $ N \sim R
 $, that is, in the conditions of the above thermodynamic limit
 (i. e. $ N/V^{1/3} $ fixed).

Then, a portion of such volume of linear size $ R_0 < R $ is
considered: at this precise point the author assumes that $ N \sim
R^3_0 $, that is, he assumes a extremely dense distribution of
particles within the volume of  size $ R_0 $. 

But for large $N$ such
distribution {\bf necessarily collapses} since the gravitational gas avoids
collapse only when $  N \sim R_0 $. In terms of the parameter $ \eta
$, the assumption $  N \sim R_0^3 $ implies 
$$ 
\eta_0  \sim N/R_0 \sim (R_0)^2 \gg 1 \; ,
$$ 
which is {\bf deep} in the collapsed phase \cite{nos1}. That is, the 
assumption $ N \sim R^3_0 $ {\bf necessarily} implies that the gas
collapses. The subsequent statements made in ref.\cite{laliena}
are direct consequences of this assumption and are not valid.

\bigskip

It is clearly true that the partition function sums over all
configurations including collapsed situations. However, these
collapsed configurations have a negligible weight for  $ \eta < \eta_0
\sim N/R_0 $. The argument of ref.\cite{laliena} applies only for  $
\eta > \eta_0 $ and only for such values of $ \eta $ collapsed states 
dominate the partition function. It must be noticed that in Nature, if
collapsed 
configurations would always dominate selfgravitating systems, then, stars,
galaxies and the interstellar medium would have collapsed since
longtime. Fortunately, this argument in ref.\cite{laliena} is not valid.

\bigskip

As stressed in refs\cite{nos1,nos2} there are {\bf two}
regimes for the selfgravitating gas in the canonical ensemble: $ \eta <
\eta_0 $ and  $ \eta > 
\eta_0 $ with $ \eta_0 = 1.51024\ldots $ for spherical geometry and
$ \eta_C \simeq 1.515  $ for cubic  geometry. For $ \eta > \eta_0 $
the selfgravitating gas {\bf do collapse} into a extremely dense phase. It
is to this {\bf collapsed} phase and {\bf only to it } that the
eqs.(30)-(33) of ref.\cite{laliena} apply. 

\bigskip

The same comments applies to the convergence of the series expansions of the
thermodynamic quantities discussed in ref.\cite{laliena} [eq.(34)  of
  ref.\cite{laliena} above]. The series in powers of $ \eta $ only
converge for $ \eta < \eta_C $. Actually, the calculation of the
radius of convergence of such series in ref.\cite{nos1,nos2} provided us 
an independent check of the numerical value of $ \eta_C $.

\bigskip

In the same token, the thermodynamic limit proposed in ref.\cite{cuba}
leads to a cataclysmic collapse for the selfgravitating gas. It is
proposed in ref.\cite{cuba} to take $ N \to \infty $ with $ V \sim
\frac{1}{N} \to 0 $. 

It is  obvious that in such limit the gas collapses since the
volume vanishes. 

More precisely, that means $ \eta \sim N^{\frac43} \to +\infty
$ deeply in the extremely dense phase. Furthermore, it is wrongly
stated in sec. 4 of  ref.\cite{cuba} 
that the the entropy is not proportional to $N$ as in refs.\cite{nos1,nos2}. 

\begin{figure}[ht!]
\epsfig{file=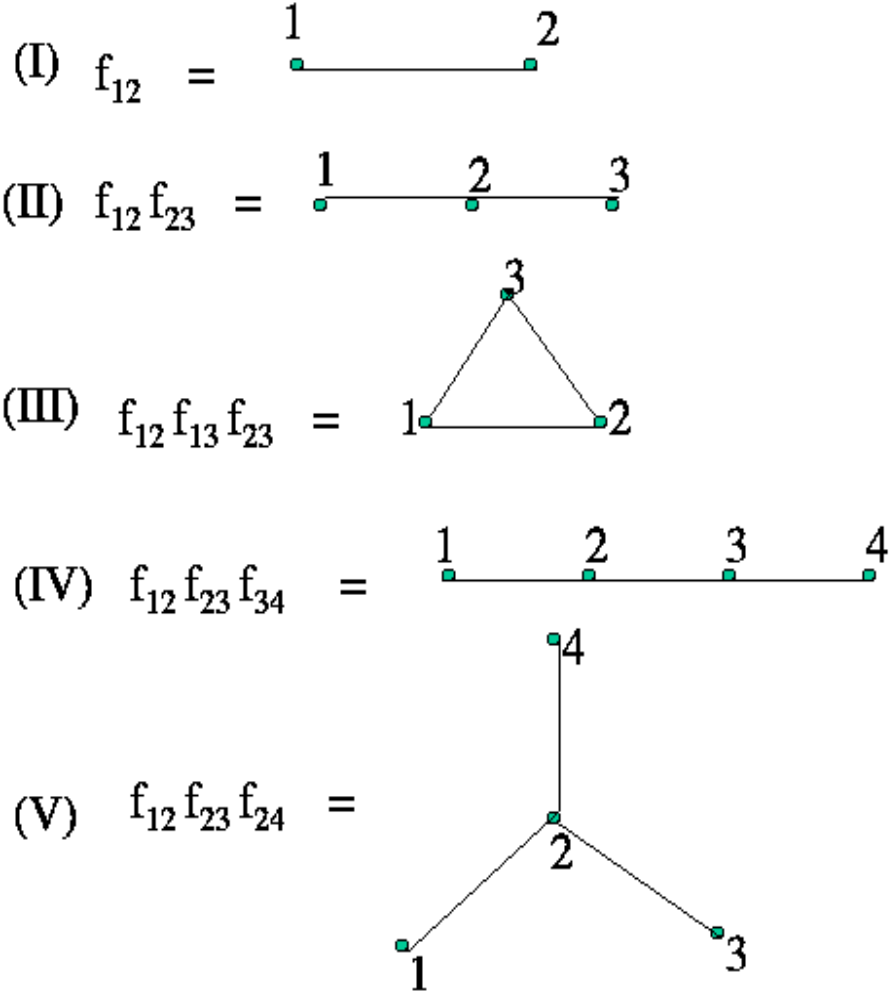,width=150mm,height=150mm}
\caption{Diagrams contributing to the cluster
  expansion. Diagram (I) gives $c_2$. Diagram (II) gives $c_3$.
  Diagrams (III), (IV) and (V) contribute to $c_4$ but  (III) is
  subdominant for large $N$. } \label{cluster}
\end{figure}

\end{document}